\documentstyle[aps,floats,amssymb,graphics]{revtex}

\begin{document}
\draft
\title{XAFS spectroscopy. I. Extracting the fine structure from the absorption
spectra}
\author {K.~V. Klementev}
\address{Moscow State Engineering Physics Institute, 115409 Kashrskoe sh. 31,
Moscow, Russia\\ e-mail: klmn@htsc.mephi.ru}
\date{\today} \maketitle

\begin{abstract}
Three independent techniques are used to separate fine structure from the
absorption spectra, the background function in which is approximated by\\ (i)
smoothing spline. We propose a new reliable criterion for determination of
smoothing parameter and the method for raising of stability with respect to
$k_{\rm min}$ variation;\\ (ii) interpolation spline with the varied
knots;\\ (iii) the line obtained from bayesian smoothing. This methods
considers various prior information and includes a natural way to determine the
errors of XAFS extraction.\\ Particular attention has been given to the
estimation of uncertainties in XAFS data. Experimental noise is shown to be
essentially smaller than the errors of the background approximation, and it is
the latter that determines the variances of structural parameters in
subsequent fitting. \end{abstract} \pacs{61.10.Ht}

\section{Introduction}
X-ray-absorption fine-structure (XAFS), $\chi$, is determined by \cite{Lytle1}:
\begin{equation}\label{chi}%%%%%%%%%%%%%%%%%%%%%%%%%%%%%%%%%%%%%%%%%%%%%%%%%%%%%%%%%%%%%
\chi(E)=[\mu(E)-\mu_0(E)]/[\mu_0(E)-\mu_b(E)],
\end{equation}%%%%%%%%%%%%%%%%%%%%%%%%%%%%%%%%%%%%%%%%%%%%%%%%%%%%%%%%%%%%%%%
where $\mu$ is the measured absorption, $\mu_0$ is the ``atomic'' absorption
due to electrons of considered atomic level, $\mu_b$ is the absorption of other
processes. Since the electronic state of an embedded atom is, in general,
different from its state in gaseous phase, $\mu_0$ is not the same as for
isolated atom and cannot be found experimentally. Therefore a demand arises for
an artificial construction of $\mu_0$.

Usually, $\mu_b$ is approximated by a Victoreen polynomial $P=aE^{-3}+bE^{-4}$
\cite{Lytle1} or by a more general polynomial $P$, coefficients of which are
found by the least squares method from $\mu(E)=P(E)$ at energies lower than the
edge.

Further, energy dependence is transformed to the photoelectron wave number
dependence: $k=\sqrt{2m_e(E-E_0)}/\hbar$, where $E_0$ is the energy of the
corresponding absorption edge. Usually, to the $E_0$ the energy at half the
step is assigned or the energy of inflection point of $\mu(E)$. In most
practical works the deviation of $E_0$ from true value, $\Delta E_0$, is one of
the fitting parameters.

The most difficult procedure in extracting of XAFS from the measured absorption
is the construction of $\mu_0$ since one cannot definitely distinguish the
environmental-born part of absorption from the atomic-like one. All methods for
determination of the post-edge background are based on the assumption of its
smoothness, and the only criterion for its validity is the absence of
low-frequncy structure in $\chi(k)\cdot k^w$, i. e. the small absolute value of
the Fourier transform (FT) $\rho(r)$ at low $r$. The review of existing
post-edge background methods and the propositions of some new is the main
purpose of the article.

Special attention must be paid to the estimation of noise and uncertainties
in XAFS data. Experimental noise is shown to be essentially smaller than the
errors of the background approximation, and it is the latter that determines the
variances of structural parameters in subsequent fitting. The corresponding
section of the present article is closely related with the next article devoted
to the determining the errors of structural parameters \cite{IIe}.

All described in the article methods for background removal, its error
estimations, and XAFS-function corrections are realized in the freeware program
{\sc viper} \cite{VIPER} which allows one to vary several parameters by hand
and watch the results simultaneously.

\section{Methods of $\mu_0$ construction}\label{methods}
\subsection{Smoothing spline}\label{smoothing}
Owing to fast algorithm and easy program realization, the approximation of
$\mu_0$ by the smoothing spline has become widespread. Let
$N+1$ experimental values of $\mu_i$ are defined on the mesh $E_i$. The
smoothing spline $\mu_0$ minimizes the functional
\begin{equation}\label{J1}%%%%%%%%%%%%%%%%%%%%%%%%%%%%%%%%%%%%%%%%%%%%%%%%%%%%%%%%%%%%%
J(\mu_0,\mu)=\int^{E_{\rm max}}_{E_{\rm min}}[\mu_0''(E)]^2\,dE+
\frac{1}{\alpha}\sum^{N}_{i=0}(\mu_{0i}-\mu_i)^2.
\end{equation}%%%%%%%%%%%%%%%%%%%%%%%%%%%%%%%%%%%%%%%%%%%%%%%%%%%%%%%%%%%%%%%
The smoothing parameter (or regularizer) $\alpha$ is the measure of compromise
between smoothness of $\mu_0$ and its deviation from $\mu$. At $\alpha=0$ the
smoothing spline exactly coincides with $\mu$, at $\alpha\to\infty$ it
degenerates to $\mu_0={\mathit const}$. Optimal regularizer should lead to
$\mu_0$ containing only low-frequency oscillations and, hence, to $\chi$
containing only structural oscillations. The formulation of a new criterion for
optimal $\alpha$ we shall consider below.

\begin{figure}[!b]\begin{center}\includegraphics*{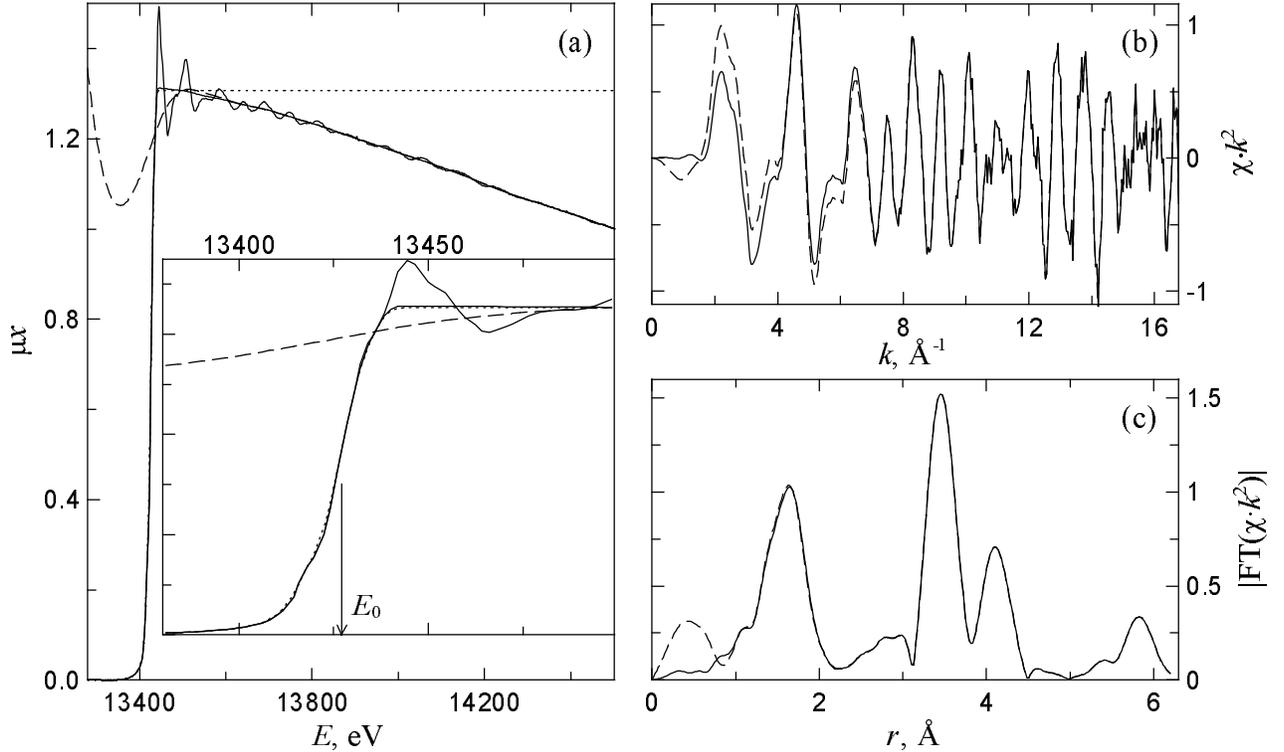}\end{center}
\caption{Extraction of XAFS from the measured absorption using the smoothing
spline. Prior function $p(E)$ for the atomic-like absorption is drawn by dots.
Solid lines --- $\mu_0(E)$, $\chi(k)\cdot k^2$, and $\rho(r)$ obtained with
use of the prior function; dashed lines --- dittos without prior function.
The regularizer $\alpha$ is the same for both cases.}
\label{mu_spl}\end{figure}

First, we address another problem, the well-known spline instability with
respect to the small variations of input parameters: number of nodes, nodal
values of the processed function, and limits on integral. In our
case the spline is most sensitive to $E_{\rm min}$ due to fast growth of $\mu$
in the edge. To raise the stability the method was put forward in {\sc viper}
program which lies in the use of a prior information specifying the shape of
$\mu_0(E)$ dependence. It is known in advance that the absorption edge without
so-called white line constitutes nearly smooth step; the white line, if
presents, is added to the step. Denote this prior function as $p(E)$. Now we
will tend the second derivative of the sought $\mu_0(E)$ not to zero (at the
specified deviation of $\mu_0$ from $\mu$) but to the second derivative of
$p(E)$. The sought $\mu_0(E)$ is now minimizes the functional
\begin{equation}\label{J2}%%%%%%%%%%%%%%%%%%%%%%%%%%%%%%%%%%%%%%%%%%%%%%%%
J^*(\mu_0,\mu)=\int^{E_{\rm max}}_{E_{\rm min}}[\mu_0''(E)-p''(E)]^2\,dE+
\frac{1}{\alpha}\sum^{N+1}_{i=0}[\mu_{0i}-\mu_i]^2.
\end{equation}%%%%%%%%%%%%%%%%%%%%%%%%%%%%%%%%%%%%%%%%%%%%%%%%%%%%%%%%%%%%
As seen, in fact there is no need to know $p(E)$ itself, its second
derivative is sufficient. The explicit presence of $p(E)$ in the following
formulas should be taken as a consequence of the technical trick applied: at
first $p(E)$ is subtracted from the data, then it is added to the found spline.

Represent the second derivatives in finite-difference approximation, introduce
$\tilde\mu_{0i}=\mu_{0i}-p_i$, and denote $\Delta_i=E_{i+1}-E_i$:
\begin{equation}\label{J3}%%%%%%%%%%%%%%%%%%%%%%%%%%%%%%%%%%%%%%%%%%%%%%%%
J^*(\mu_0,\mu)=\sum^N_{i=1}[\tilde\mu_{0i-1}\Delta_{i-1}^{-1}-
\tilde\mu_{0i}(\Delta_{i-1}^{-1}+\Delta_i^{-1})+\tilde\mu_{0i+1}\Delta_i^{-1}]^2+
\frac{1}{\alpha}\sum^{N+1}_{i=0}[\tilde\mu_{0i}-(\mu_i-p_i)]^2=
J(\tilde\mu_0,\mu_i-p_i).
\end{equation}%%%%%%%%%%%%%%%%%%%%%%%%%%%%%%%%%%%%%%%%%%%%%%%%%%%%%%%%%%%%
Thus, the problem is reduced to the preceding one in which instead of initial
data $\mu_i$ the difference $\mu_i-p_i$ is appeared. The sought $\mu_0$ is
found from the smooth $\tilde\mu_0$ as $\mu_{0i}=\tilde\mu_{0i}+p_i$.
In Fig.~\ref{mu_spl} is shown an example of the atomic-like absorption
approximation by the smoothing spline with and without the use of prior
function\footnote{Here and hereafter for examples is used the spectrum at Bi
$L_3$ absorption edge in Ba$_{0.6}$K$_{0.4}$BiO$_3$ at 50\,K recorded in
transmission mode at D-21 line (XAS-13) of DCI (LURE,Orsay, France) at positron
beam energy 1.85 GeV and the average current $\sim250$\,mA.  Energy step ---
1\,eV, counting time --- 1\,s.  Energy resolution of the double-crystal Si
[311] monochromator (detuned to reject 50\% of the incident signal in order to
minimise harmonic contamination) with a 0.4\,mm slit was about 2--3\,eV at
13\,keV.}. Energy $E_0$ was determined at half the step height. Here, we
constructed $p(E)$ in the following manner. Found the average value $\bar\mu$
of $\mu(E)$ in region $20\le E\le70$\,eV above the absorptance maximum. Moving
from the beginning of spectrum, assign $p=\mu$ until $\mu>\bar\mu$, further
$p=\bar\mu$. Then $p(E)$ was smoothed 5 times on 3 points. To perform the
Fourier transform, $\chi(k)k^2$ was brought into the uniform scale with
$\delta k=0.03$\,\r{A}$^{-1}$ and multiplied by a Kaiser-Bessel window with
parameter $A=1.5$. As seen, the use of $p(E)$ has led to disappearance of the
spurious peak on the absolute value of FT at $r\sim0.5$\,\r{A}.

So far we have considered the atomic-like absorption $\mu_0$ to be a
smooth function with no peculiarities. However, in some spectra $\mu_0$ itself
has a fine structure \cite{Rehr1,Filipponi2} originating from resonance
scattering within absorbing atom or from multi-electron transitions. If in
these cases, based on theoretical calculations, experimental information, or
empirical considerations, one can nearly indicate the location of
peculiarities, their width and weight relatively to the step height, then one
would readily construct the prior function $p(E)$ and find the correct $\mu_0$.
Instead of constant value above absorption edge, the prior function would have
corresponding valleys and/or peaks.

Let us now define the criterion for determination of smoothing parameter.
An attempt to solve the problem was made in Ref.~\cite{Cook1}, where
the requirement was proposed: $H_R-H_N\ge0.05H_M$, where $H_R$ is the average
value of the weighted Fourier transform magnitude between 0 and 0.25\,\r{A},
$H_M$ is the maximum value in the transform magnitude between 1 and 5\,\r{A},
$H_N$ is the average value of the transform magnitude between 9 and 10\,\r{A}
attributed to the noise. Obviously, that this criterion cannot pretend to the
generality since depends on the weighting (op. cit., $k^3$) and the relative
contribution of noise and the first coordination shell into spectra.

\begin{figure}[!t]
\begin{minipage}[c]{0.6\hsize}\includegraphics*{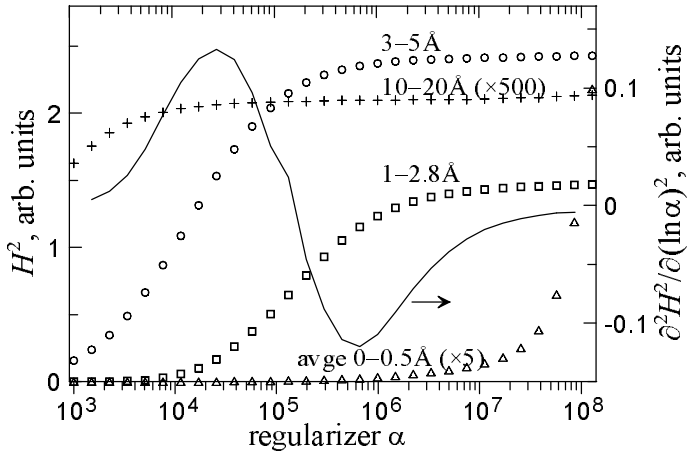}\end{minipage}
\begin{minipage}[c]{0.35\hsize}\caption{The $\rho(r)$ peak heihts squared,
$H^2$, maximal in the indicated areas and average over the range
$0<r<0.5$\,\r{A}, as functions of $\alpha$. To the right axis relates the
second derivative of the first peak height squared with respect to
$\ln\alpha$.} \label{regular}\end{minipage}\end{figure}

In the program {\sc viper} we have proposed another approach to the problem
based on the consideration of heights of FT peaks as functions of regularizer
$\alpha$ (see Fig.~\ref{regular}). On increasing $\alpha$ from zero, $\mu_0$
starts to deviate from the experimental absorption $\mu$, $\rho(r)$ is growing
and then saturates, the peaks at larger $r$ being saturated earlier. Clearly,
that $\alpha$ should be determined by the first peak height since it is the
last to saturate. Define the start of saturation on the minimum of second
derivative of the first peak squared with respect to $\ln\alpha$. Declare
the value of $\alpha$ in the minimum to be optimal. It is seen that the
increase of $\alpha$ from the optimal leads to unwanted rapid growth of $\rho$
at low $r$.

In the example in Fig.~\ref{mu_spl} the regularizer is optimized following our
new criterion.

Unfortunately, the method of smoothing spline does not include any approach to
the estimations of uncertainties in the $\mu_0$ obtained, in contrast to the
following two methods.

\subsection{Interpolation spline drawn through the varied knots}
\label{knots}
The method was put forward in Ref.~\cite{Newville1}. $N$ knots are equally
spaced in $k$ space, through them an interpolation spline is drawn. The
ordinates of the knots are varied to minimize $\rho$ or $|\rho-\rho_{\rm st}|$
in the chosen low-$r$ region $0\le r\le r_0$, where $\rho_{\rm st}$ is the
absolute value of the FT of a ``standard'' $\chi_{\rm st}(k)\cdot k^w$,
calculated or experimental. The number of knots must not exceed the value
$N_{\rm max}=2r_0\Delta k/\pi+1$, \cite{Stern1} where $\Delta k$ is the $k$
range of useful data. In the Ref.~\cite{Newville1} was asserted that one need
to know the ``standard'' $\chi_{\rm st}(k)\cdot k^w$ merely approximately
since it used only to get an estimate of the leakage from the first shell to
the region minimized. The strange thing is that having omitted the question on
the accuracy of found knots (as we show below, rather poor), the authors of the
cited work made a fine comparison between several theoretical models for
$\chi(k)$ calculations.

In Fig.~\ref{mu_spl} is shown an example of the method application. Ordinates
of the 13 knots ($N_{\rm max}=13.2$) were varied to minimize the difference
$\rho-\rho_{\rm st}$ at $0\le r\le1.05$\,\r{A}. The function
$\chi_{\rm st}(k)$ was calculated using {\sc feff6} program \cite{FEFF} (as was
pointed above, a crude estimate is sufficient, so details omitted). In the
minimized region the $\rho(r)$ is somewhat better than that obtained by the
previous method. However, at $k>15$\,\r{A}$^{-1}$ one can distinguish the
obviously wrong behavior of $\chi(k)\cdot k^2$, and the first peak on
$\rho(r)$ becomes quite distorted.
\begin{figure}[!b]\begin{center}\includegraphics*{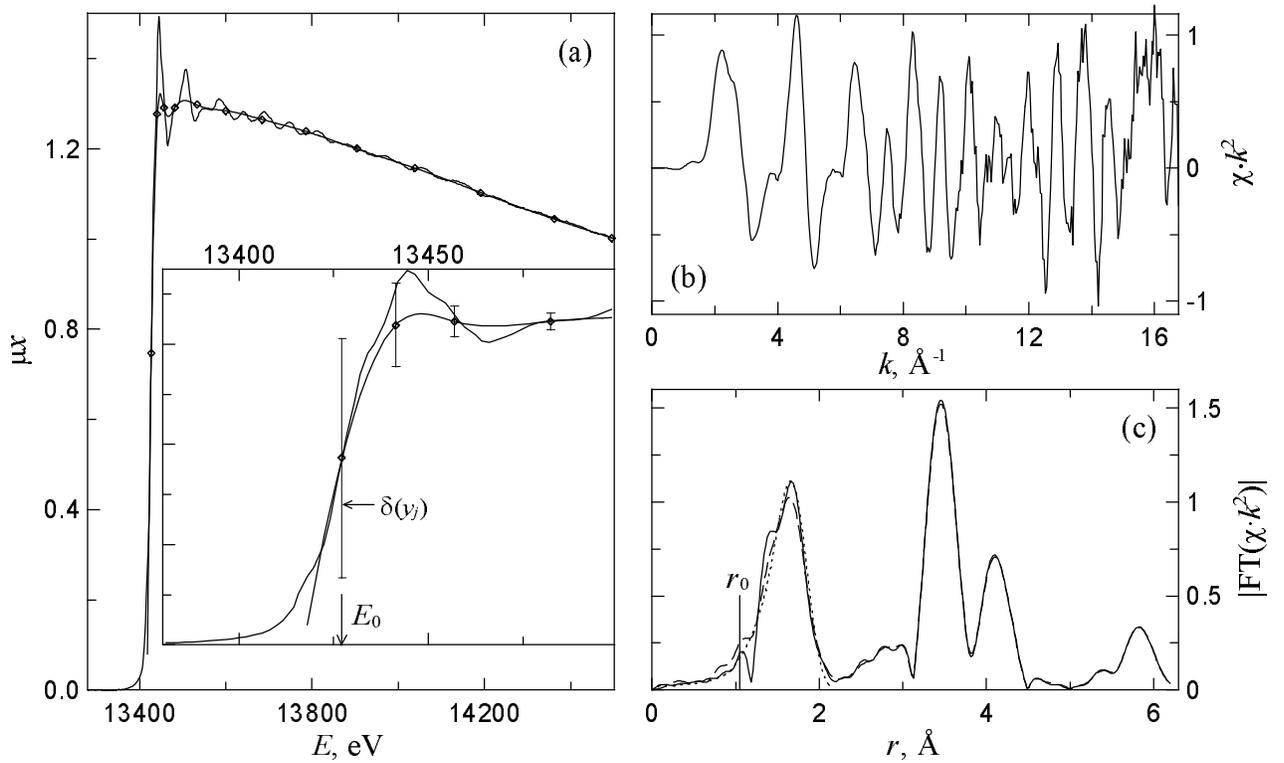}\end{center}
\caption{Extraction of XAFS from the measured absorption using the
interpolation spline through the knots with varied ordinates. On (c) the
Fourier transforms are shown for sought $\chi(k)\cdot k^2$ (solid),
``standard'' (dots), and obtained by the previous method (dashed).}
\label{mu_knt}\end{figure}

\begin{figure}[!t]
\begin{minipage}[c]{0.55\hsize}\includegraphics*{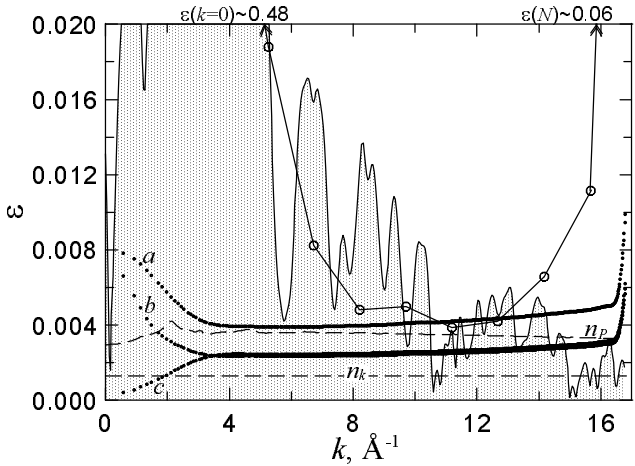}\end{minipage}
\begin{minipage}[c]{0.4\hsize}\caption{Errors of $\chi(k)$ extraction.
Solid line with open circles --- by the method of interpolation spline drawn
through the varied knots. Dots --- by the method of bayesian smoothing without
($a$) and with ($b$ ¨ $c$) prior information specifying the second derivative.
Besides, ($c$) uses additional information that $\mu_0(E)$ passes through a
point immediately before $E_0$. Solid line with filling --- the envelope of
$\chi(k)$ (not weighted). Dashed lines --- the noise estimates from FT
($n_k$) and from Poisson counting statistics ($n_P$) (see
Sec.~\protect\ref{limits}).} \label{chi_epsi}\end{minipage}\end{figure}

Consider now the problem of the accuracy of knot positions $y_j$,
$j=1,\ldots,N$ in fitting $\rho(r)$ to $\rho_{\rm st}(r)$.
As a figure of merit, the $\chi^2$-statistics appears:
\begin{equation}\label{K1}%%%%%%%%%%%%%%%%%%%%%%%%%%%%%%%%%%%%%%%%%%%%%%%%%%
\chi^2=\frac{N_{\rm max}}{M}\sum_{m=1}^{M}\frac{[\rho(r_m)-
\rho_{\rm st}(r_m)]^2}{\sigma_m^2},
\end{equation}%%%%%%%%%%%%%%%%%%%%%%%%%%%%%%%%%%%%%%%%%%%%%%%%%%%%%%%%%%%%%%%
where $\sigma_m$ are the errors of $\rho(r_m)$. It can be shown (detailed
analysis see in the next article \cite{IIe}) that under the assumption of
uncorrelated knot positions, the mean-square deviation of $y_j$ from the
obtained through the fit optimal value $\hat y_j$ equals
$\delta(y_j)=(\frac{1}{2}\partial^2\chi^2/\partial y_j^2)^{-1/2}$, where the
partial derivatives are calculated in the fitting procedure at the minimum.
$\sigma_m$ are assumed to be constant and equal to the root-mean-square
average of $\rho(r)$ between 15 and 25\,\r{A}, where solely the noise is
present. The errors $\varepsilon_j=\delta(y_j)/[\mu_0(E_j)-\mu_b(E_j)]$
found under such assumptions are shown in Fig.~\ref{chi_epsi} as open circles
with the solid line. Notice, that the ussumption that the knot positions are
not correlated gives quite optimistic $\varepsilon_j$. Actually, several first
knot positions appear to be highly correlated; the proper taking into account
of the correlations (here we do not present these calculations) raises
$\varepsilon_j$ at the least as twice. But even these underestimated
$\varepsilon_j$ are appreciably larger than those given by the following
method.

\subsection{Bayesian smooth curve}\label{bayesian}
Ideologically similar to the smoothing spline method is the method of bayesian
smoothing (see Appendix on p.~\pageref{BayesianMethod}) proposed in
the program {\sc viper}. This method also finds the regularized function
$\mu_0$, the regularizer $\alpha$ is the measure of compromise between
smoothness of $\mu_0$ and its deviation from $\mu$. In comparison with
smoothing spline method, this method has some advantages. (i) Various prior
information on $\mu_0$ can be considered. (ii) In this method the
posterior distributions of all $\mu_{0j}$ are sought for. From those
distributions one can find not only average values but also any desirable
momenta, which appears to be an additional difficulty for other methods. (iii)
In the framework of the method it is possible also to deconvolute $\mu$ with
the monochromator rocking curve. The weakness of the method is its low speed
(comparing with method \ref{smoothing}, not with \ref{knots}!). On a modern PC
the curve drawn through $N\sim500$ points is smoothed for a few minutes.
\begin{figure}[!t]
\begin{center}\includegraphics*{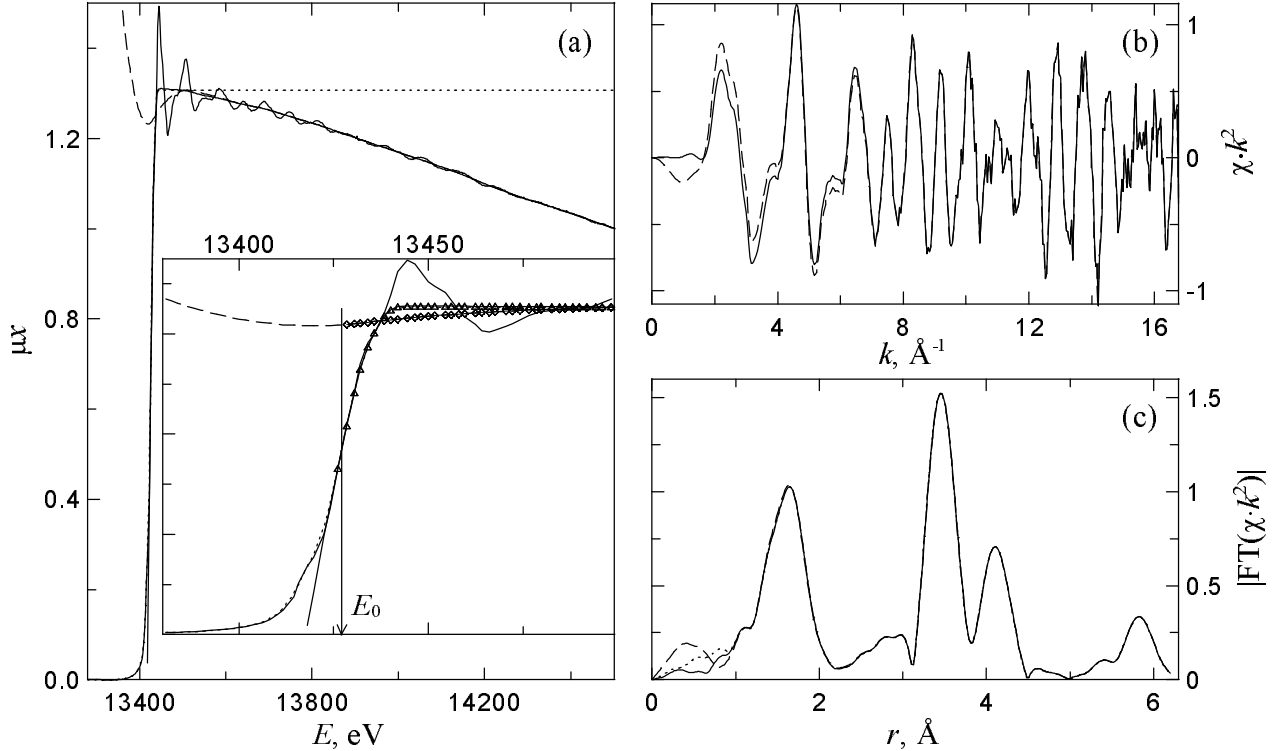}\end{center}
\caption{Extraction of XAFS from the measured absorption using the
bayesian smoothing. Prior function $p(E)$ for the atomic-like absorption is
drawn by dots. Solid lines --- $\mu_0(E)$, $\chi(k)\cdot k^2$, and $\rho(r)$
obtained with the use of the prior function; dashed lines --- dittos without
prior function. The dot line on (c) is obtained without additional requirement
for $\mu_0(E)$ to pass through a point immediately before $E_0$. The
regularizer $\alpha$ is the same for all cases and equals to the optimal one
found for the smoothing spline.} \label{mu_bayes}\end{figure}

In Fig.~\ref{mu_bayes} the bayesian smoothing was done on the mesh of 536
experimental points above $E_0$, without and with the prior function (its
construction is described in Sec.~\ref{smoothing}). Besides, in the last case
another information was used: the atomic-like absorption must coincide with the
total absorption (minus pre-edge background) at energies $E<E_0$. Therefore, we
demanded from the bayesian curve to pass through a point nearest (at left) to
$E_0$. The values $\bar\mu_{0j}$ and $\delta^2(\mu_{0j})$ were found by
formulas (\ref{B26_3}) and (\ref{B26_5}). Since the smoothed values do not lie
within the limits of $\pm\delta(\mu_{0j})$ from $\mu_j$, we did not look for the
most probable smoothness (see.~Appendix), instead we considered the regularizer
to be known and equal to the optimal one found in the method \ref{smoothing}.
The introduction of the prior information has significantly diminished the
errors of $\chi(k)$ extraction (see dotted curves in Fig.~\ref{chi_epsi}) which
were defined as $\varepsilon_j=\delta(\mu_{0j})/(\bar\mu_{0j}-\mu_{bj})$. This
is quite natural: any decrease of our ignorance about $\mu_0$ should narrow the
posterior distribution of $\mu_{0j}$ for all $j$. Of course, this concerns the
experimental information as well: errors $\varepsilon_j$ are the less the more
measured points $N$ the spectrum has. Comparing Fig.~\ref{mu_spl}(á) and
Fig.~\ref{mu_bayes}(c), it is seen practically perfect coincidence of the
results of bayesian smoothing and smoothing spline. From this one can assume
the equality of the errors which both methods give.

Could we take into account possible systematic errors in the framework of
the method? Yes, if we have the information on their nature and are able to
translate it into the mathematics language; such a translation might be rather
non-trivial. In any case, now we have the tool to extract from the prior and
experimental information not only the sought values but their errors as well.

\subsection{Other methods}\label{OtherMethods}
Consider briefly the methods for $\mu_0$ construction not included into the
{\sc viper} program.

A rich variety of computer programs for XAS spectra processing is collected
on the International XAFS Society Web-site \cite{Catalog}. The vast majority of
them use as an approximation for the atomic-like absorption a smoothing spline
or more general piecewise-polynomial representation. For example, in the method
of Ref.~\cite{Kuzmin1}, the construction of $\mu_0$ is divided into several
stages: $\mu_0$ is approximated by a low-degree polynomial, obtained $\chi(k)$
is multiplied by $k^w$, additional $\mu_0'$ is drawn again as a low-degree
polynomial and subtracted, a smoothing spline then approximates one more
additional $\mu_0''$. The sum of all $\mu_0$'s gives the total atomic-like
background. The necessity of the preliminary stages was not discussed op. cit.,
however, clearly it was caused by the instability of spline with respect to the
small variations of input parameters. And the point is not that the preliminary
stages make the process stable, but that for each specific spectrum, auxiliary
parameters (degrees of polynomials) could provide an acceptable construction of
the atomic-like background. Above (in Sec.~\ref{smoothing}) we proposed the way
to rise the stability of spline making the preliminary stages to be redundant.

In Ref.~\cite{Bridges1} an iterative approach to ``atomic background'' removal
was developed. First a spline is used to obtain a rough estimate of the
background; this alone is enough to have a reliable $\chi$ at
$k>5-6$\,\r{A}$^{-1}$. Over that range the $\chi$ obtained is fitted to the
theoretical $\chi_{\rm th}$ in $r$-space. The resulting fit parameters are
used then to generate $\chi_{\rm th}(k)$ that extends down to low $k$. This
function is transformed back into $e$-space and $\mu_0$ is obtained as
$\mu_0=\mu/(\chi_{\rm th}+1)$ that need be a little smoothed or fitted by an
additional spline. Since the logic of reasoning was inversed: not ``find
$\mu_0$ to find $\chi$,'' but ``find $\chi$ to find $\mu_0$,'' the method is
suited for the quest of peculiarities on $\mu_0$ curve, not for structural
XAFS-researches. Besides, the range of accuracy of the model appears to be
unknown in principle: all, that is not described by the model, is included in
$\mu_0$; the errors of the background approximation are also undefined.

In the old work \cite{Boland1} for the determination of the background
absorption $\mu_0$ was considered the damping of the XAFS amplitude resulting
from the measurements with low resolutions (with a large slit width). The
superimposition of two spectra measured with different energy resolutions gives
the intersection points, the part of which belong to the $\mu_0$. Then through
the obtained nodal points a smoothing spline is drawn. As the authors of
Ref.~\cite{Boland1} noted, the measurements of the spectra with worsened
resolution are not necessary; the spectra could be damped by the convolution
with a ``rocking curve,'' approximated by a Gaussian function. Of course, the
method is correctly works only with a small variation of the Gaussian curve
width since for the large width not only the XAFS amplitude is damped but the
very edge is washed out. Because of this only the extended part of a spectrum
could be reliably determined.

The damping of the XAFS amplitude can be due to other reasons. For instance,
as was pointed in Ref.~\cite{Boland1}, the nodal points may be obtained from
the variable-temperature study. This idea was realized in Ref.~\cite{Stern2}
and is more sound since the atomic-like background is really independent of
temperature and with temperature the XAFS amplitude is changed, not the shape of
the edge. But for all that it is important that the phase difference between
XAFS of different temperatures was negligible, which is true only for low wave
numbers. Unfortunately, the method is suitable only for some particular cases
(to say nothing of need for measured temperature series of spectra). Op. cit.
it was demonstrated for the x-ray-absorption data for the $L_3$ edge of solid
Pb. In those spectra the first crossing of $\mu$ and $\mu_0$ occurs already at
$\sim15$\,eV above edge. In our sample spectra the first crossing occurs only
at $\sim30$\,eV, which allows one to find at most 2--3 points and the first of
them being situated at $k\gtrsim2.5$\,\r{A}$^{-1}$.

\begin{figure}[!t]
\begin{minipage}[c]{0.55\hsize}\includegraphics*{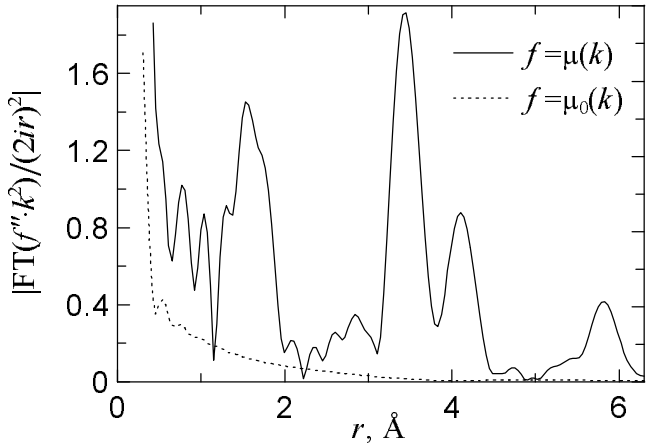}\end{minipage}
\begin{minipage}[c]{0.4\hsize}\caption{On the method of
Ref.~\protect\cite{Hu1}. Solid curve is the FT of ``unnormalized
$\chi(k)\cdot k^2$'', dotted curve is the contribution from the atomic-like
absorption.}
\label{hu_met}\end{minipage}\end{figure}

An interesting approach to the problem of $\mu_0$ determination was reported in
Ref.~\cite{Hu1}. It is based on the simple identity that relates the FT of some
function with the FT of its $n$-th derivative:
\begin{equation}\label{Hu}%%%%%%%%%%%%%%%%%%%%%%%%%%%%%%%%%%%%%%%%%%%%%%%%%%
{\rm FT}[f^{(n)}(k)]=(2ir)^n{\rm FT}[f(k)],
\end{equation}%%%%%%%%%%%%%%%%%%%%%%%%%%%%%%%%%%%%%%%%%%%%%%%%%%%%%%%%%%%%%%%
where the conjugate variables are $k$ and $2r$. Since the atomic-like
background is smooth enough, the higher derivatives $\mu^{(n)}(k)$ ($n\ge2$)
are oscillatory near zero. Performing the FT of $\mu^{(n)}(k)\cdot k^w$ and
using Eq.~(\ref{Hu}), one obtains the FT of unnormalized would-be
$\chi(k)\cdot k^w$ (see Fig.~\ref{hu_met}). Op. cit. the low-$r$ part (which in our example
is $0\le r\lesssim1.1$) was cut off, and then the back FT was done. As a
result, one has the unnormalized $\chi(k)\cdot k^w$ and, having subtracted it
from the $\mu(k)$, the atomic-like background on which some peculiarities due
to multi-electron excitations can be distinguished. Like the method of
Ref.~\cite{Bridges1}, this method is suited for the quest of peculiarities on
the $\mu_0$ curve, not for structural XAFS-researches because of evident
distortion of the first peak on the FT by the contribution from the
atomic-like background. To illustrate this assertion, in Fig.~\ref{hu_met} we
show the FT of the second derivative of the $\mu_0(k)$ that was found by the
present method. As seen, this contribution is not as small.

If the electronic states of an absorbing atom in gaseous phase and in the
compound of interest may be considered as equivalent, $\mu_0$ can be set equal
to the measured absorption in gas, as was done in Ref~\cite{DiChicco1} for
solid, liquid, and gaseous Kr. Some differences in energy positions and
relative weights of double-electron excitation channels were taken into
account by a model using simple empirical functions which were transferred then
to the spectra of liquid and solid Kr. Notice that the proposed in the present
paper prior function for the methods of smoothing spline and bayesian smoothing
can include additional items corresponding to the multi-electron contributions.

\section{Errors in $\mu_0$ constructing, noise, and choice of limits
\lowercase{$k_{\rm min}$} ¨ \lowercase{$k_{\rm max}$}}\label{limits}
For what we need to know the errors of XAFS-function extraction? First, without
knowing of these values one cannot in principle aim at their minimization.
Second, they are used in the definition of $\chi^2$-statistics in the fitting
problems; their underestimation is a source of unjustified optimistic errors of
fitting parameters. Third, along with analysis of the noise, the errors of
$\mu_0$ construction allow us to choose the limits of reliable EXAFS signal,
$k_{\rm min}$ and $k_{\rm max}$.

Unfortunately, the issue of quality of XAFS extraction from the measured
absorption has not been addressed properly. We see several reasons for that.
On the one hand, not having a correctly developed approach to the estimation of
the errors of final results (interatomic distances, Debye-Waller factors etc.
found via fitting), the errors of EXAFS extraction are useful. On the other
hand, only a few methods include approaches to their estimations.

Easily one can compare the errors of different methods (see
Fig.~\ref{chi_epsi}) and then choose the most reliable one. The problem of
plausible limitations on the absolute value of the errors is more difficult.
Define ``signal'' as the envelope of $\chi(k)$ (solid line with gray filling in
Fig.~\ref{chi_epsi}). It is quite reasonable to demand that the errors of
$\mu_0$ construction were less than XAFS signal. For the method of the
interpolation spline drawn through the varied knots to meet this requirement
leads to the restriction on the photoelectron wave numbers:
$2\lesssim k\lesssim14$\,\r{A}$^{-1}$. For the bayesian curve $a$ this range is
$0\le k\lesssim14$\,\r{A}$^{-1}$, for the bayesian curves $b$ and $c$ this
range is wider: $0\le k\lesssim16$\,\r{A}$^{-1}$.

Another factor that limitates the spectrum length is the presence of noise.
To determine the noise is a straightforward task for $r$-space, where XAFS
signals at high $r$ have clearly noise character. By Parseval's identity the
noise in $r$-space is related with the noise in $k$-space \cite{Newville2}:
\begin{eqnarray}\label{Noise1}%%%%%%%%%%%%%%%%%%%%%%%%%%%%%%%%%%%%%%%%%%%%%%
\int_{k_{\rm min}}^{k_{\rm max}}|n_kk^w|^2dk=2\int_{0}^{\pi/2dk}|n_r|^2dr.
\end{eqnarray}%%%%%%%%%%%%%%%%%%%%%%%%%%%%%%%%%%%%%%%%%%%%%%%%%%%%%%%%%%%%%%
Substitute the mean value over the range $15<r<25$\,\r{A} of the FT magnitude
squared for $|n_r|^2$. Then
\begin{eqnarray}\label{Noise2}%%%%%%%%%%%%%%%%%%%%%%%%%%%%%%%%%%%%%%%%%%%%%%
n_k^2=\langle|n_r^2|\rangle\frac{\pi}{dk}\
\frac{2w+1}{k_{\rm max}^{2w+1}-k_{\rm min}^{2w+1}}.
\end{eqnarray}%%%%%%%%%%%%%%%%%%%%%%%%%%%%%%%%%%%%%%%%%%%%%%%%%%%%%%%%%%%%%%
As seen from the formula, $n_k$ depends on $dk$, the size of evenly-spaced
$k$-grid. Although above we already have used the Fourier transform, the
question of choice of $dk$ was not raised yet. The algorithm of fast FT needs
the transformed function to be set on a uniform grid. Having chosen a small
$dk$, we artificially obtain the large number of ``experimental'' values.
Naturally, this trick would not give more information than we have, and the
errors $n_k$ must be large at the small $dk$. In our example the choice of $dk$
(0.03\,\r{A}$^{-1}$) was based on the equality of numbers of experimental points
and the nodes of the grid. The signal-to-noise ratio obtained is greater than
unity for all the spectrum (see Fig.~\ref{chi_epsi}). There was no doubt in
that: the signal is visually distinguished even for the very extended end of
the spectrum (see Fig.~\ref{mu_spl}(b) and Fig.~\ref{mu_bayes}(b)).

The noise can be estimated based on the bayesian considerations \cite{Gull1}.
Let after measurements we have the values of counts from the solid-state or
gas-filled detectors and let there is a positive real number $\lambda$ such
that the probability that a single count occurs in the time interval $dt$ is
\begin{eqnarray}\label{Poisson0}%%%%%%%%%%%%%%%%%%%%%%%%%%%%%%%%%%%%%%%%%%%%%%
P(1|\lambda)=\lambda dt.
\end{eqnarray}%%%%%%%%%%%%%%%%%%%%%%%%%%%%%%%%%%%%%%%%%%%%%%%%%%%%%%%%%%%%%%
It can be shown \cite{Jaynes1} that merely from this assumption follows that
the counts obey the Poisson distribution law:
\begin{eqnarray}\label{Poisson1}%%%%%%%%%%%%%%%%%%%%%%%%%%%%%%%%%%%%%%%%%%%%%%
P(N|\lambda,T)=\frac{(\lambda T)^N\exp(-\lambda T)}{N!},
\end{eqnarray}%%%%%%%%%%%%%%%%%%%%%%%%%%%%%%%%%%%%%%%%%%%%%%%%%%%%%%%%%%%%%%
where $T$ is the sampling time. The problem is to find the intensity $\lambda$
and its variance. Using Bayes theorem and introducing prior probabilities
$P(N)=1/N$ and $P(\lambda)=1/\lambda$ \cite{Jeffreys1}, one obtains:
\begin{eqnarray}\label{Poisson2}%%%%%%%%%%%%%%%%%%%%%%%%%%%%%%%%%%%%%%%%%%%%%%
P(\lambda|N,T)=\frac{P(N|\lambda,T)P(\lambda)}{P(N)}=
\frac{T(\lambda T)^{N-1}\exp(-\lambda T)}{(N-1)!},
\end{eqnarray}%%%%%%%%%%%%%%%%%%%%%%%%%%%%%%%%%%%%%%%%%%%%%%%%%%%%%%%%%%%%%%
that is after measurement the variate $2T\lambda$ follows the
$\chi^2$-distribution with $2N$ degrees of freedom. It is easy to find that
$\bar\lambda=N/T$, $\overline{\lambda^2}=N(N+1)/T^2$, and the variance of
intensity is $\delta\lambda=\sqrt{N}/T$.

Denote counts from detectors measuring $i_0$ and $i_1$ as $I_0$ and $I_1$.
By definition the variate $\xi=\frac{i_0/2I_0}{i_1/2I_1}$ follows Fisher's
$F$-distribution with $(2I_0,2I_1)$ degrees of freedom. Its expected value and
variance are known:
$\bar\xi=I_1/(I_1-1)$, $\delta^2\xi=I_1^2(I_0+I_1-1)/((I_1-1)^2(I_1-2)I_0)$,
from where we find for the absorption in the fluorescence mode
($\mu x=i_0/i_1$):
\begin{eqnarray}\label{Poisson3}%%%%%%%%%%%%%%%%%%%%%%%%%%%%%%%%%%%%%%%%%%%%%%
\overline{i_0/i_1}=\frac{I_0}{I_1-1},\qquad\delta^2(i_0/i_1)=
\frac{I_0(I_0+I_1-1)}{(I_1-1)^2(I_1-2)}.
\end{eqnarray}%%%%%%%%%%%%%%%%%%%%%%%%%%%%%%%%%%%%%%%%%%%%%%%%%%%%%%%%%%%%%%
Further, the variate $\eta=\frac{1}{2}\ln\xi$ follows $z$-distribution
(Fisher's distribution of variance ratio) with $(2I_0,2I_1)$ degrees of
freedom. Its expected value and variance are known: $\bar\eta=0$,
$\delta^2\eta=\frac{1}{4}(I_1+I_1)/(I_0I_1)$, from where we find for the
absorption in the transmission mode ($\mu x=\ln(i_0/i_1)$):
\begin{eqnarray}\label{Poisson4}%%%%%%%%%%%%%%%%%%%%%%%%%%%%%%%%%%%%%%%%%%%%%%
\overline{\ln(i_0/i_1)}=\ln\frac{I_0}{I_1},\qquad\delta^2\ln(i_0/i_1)=
\frac{1}{I_0}+\frac{1}{I_1}.
\end{eqnarray}%%%%%%%%%%%%%%%%%%%%%%%%%%%%%%%%%%%%%%%%%%%%%%%%%%%%%%%%%%%%%%
The noise of XAFS-function is
\begin{eqnarray}\label{Poisson5}%%%%%%%%%%%%%%%%%%%%%%%%%%%%%%%%%%%%%%%%%%%%%%
n_P=\frac{\delta\mu}{\mu_0-\mu_b}=
\left(\frac{1}{I_0}+\frac{1}{I_1}\right)^{1/2}\frac{1}{\mu_0-\mu_b}.
\end{eqnarray}%%%%%%%%%%%%%%%%%%%%%%%%%%%%%%%%%%%%%%%%%%%%%%%%%%%%%%%%%%%%%%

In our example this noise at $k\gtrsim15$\,\r{A}$^{-1}$ becomes greater than
signal (see Fig.~\ref{chi_epsi}). What is the reason for such significant
difference between the really present noise $n_k$ and its statistical estimate
$n_P$? Of course, the reason is in the false premise (\ref{Poisson0}). In
practice this condition is realized as: $P(c|\lambda)=\lambda dt$. For example,
the photocurrent in an ion-chamber depends on gas pressure, potential applied
etc.; these dependencies are contained in $c$. In other words, the
amplification path works in such a way that one photon gives birth to $c$
counts. There is no difficulty in writing the posterior distribution for the
generalized premise:
\begin{eqnarray}\label{Poisson6}%%%%%%%%%%%%%%%%%%%%%%%%%%%%%%%%%%%%%%%%%%%%%%
P(\lambda|N,T)=\frac{T(\lambda T)^{\frac{N}{c}-1}\exp(-\lambda T)}{\Gamma(N/c)},
\end{eqnarray}%%%%%%%%%%%%%%%%%%%%%%%%%%%%%%%%%%%%%%%%%%%%%%%%%%%%%%%%%%%%%%
with $\bar\lambda=N/(cT)$ and $\delta\lambda=\sqrt{N/c\phantom\,}/T$. Thus,
having unknown $c$ (and implicitly assigning $c=1$), we got wrong variances for
$i_0/i_1$ and $\ln(i_0/i_1)$. Unfortunately, in the most of real experiments
the association between the probability of a single count event and the
radiation intensity (via $c$) is unknown. In spite of this, the Poisson counting
statistics is traditionally used for a long time. For example, in
Ref.~\cite{Lee2} signal-to-noise ratios are evaluated (assuming $c=1$) for the
different detection schemes.

Practically all programs for XAFS spectra processing \cite{Catalog} to estimate
the noise use the Fourier analysis. But then it is the noise that they use as
uncertainties $\varepsilon_i$ of $\chi(k)$ determination in definition of
$\chi^2$-statistics:
\begin{eqnarray}\label{Poisson7}%%%%%%%%%%%%%%%%%%%%%%%%%%%%%%%%%%%%%%%%%%%%
\chi^2=\frac{N_{\rm max}}{M}\sum_{i=1}^{M}
\frac{[(\chi_{\rm exp})_i-(\chi_{\rm mod})_i]^2}{\varepsilon_i^2}.
\end{eqnarray}%%%%%%%%%%%%%%%%%%%%%%%%%%%%%%%%%%%%%%%%%%%%%%%%%%%%%%%%%%%%%%
It would be more correct to consider as $\varepsilon_i$ the {\em larger} from
the two: the noise and the errors of the construction of $\mu_0$. In our case
(and as a rule) the latter are essentially greater (especially in the method
\ref{knots}) than the noise. In the following paper \cite{IIe} we shall show
how the understated $\varepsilon_i$ lead to optimistic errors of structural
parameters.

\section{XAFS-function correction}\label{corrections}
Because of one reason or another the experimental XAFS might be distorted.
Consider some of them.

(i) Let the counts ($I$) from detectors are associated with the intensities
($i$) as $i_0=\varkappa_0I_0$ and $i_1=\varkappa_1I_1$. Then the absorption (in
the transmission mode) equals:
\begin{eqnarray}\label{correct1}%%%%%%%%%%%%%%%%%%%%%%%%%%%%%%%%%%%%%%%%%%%%
\mu x=\ln(i_0/i_1)=\ln(I_0/I_1)+\ln(\varkappa_0/\varkappa_1).
\end{eqnarray}%%%%%%%%%%%%%%%%%%%%%%%%%%%%%%%%%%%%%%%%%%%%%%%%%%%%%%%%%%%%%%
The second term is a slightly varied function of energy and can be taken into
account in independent experiments. Such a distortion appears, for instance, if
the absorptance of the gas in ion-chamber detectors depends on energy.

(ii) If some part of incident radiation is not attenuated in the sample as much
as expected (due to the pinholes in the sample, harmonics in the incoming beam
etc.), that is $i_0=\varkappa_0I_0+b$, then the real absorption is connected
with the measured $I_0$ and $I_1$ in a complicated way. In Ref.~\cite{Stern3}
the possible decrease of XAFS amplitude shown to be essential even at low
$b/(\varkappa_0I_0)$ but thick samples. At known ratio $b/(\varkappa_0I_0)$,
the correcting factor can be easily obtained.

(iii) In the fluorescence mode, due to absorptance of the fluorescent signal in
the sample itself XAFS spectra strongly depend on the detection geometry. In
Ref.~\cite{Troeger1} the correcting functions are found explicitly.

(iv) The problem of glitches is widespread in the XAFS analysis. The glitches
are due to multiple Bragg reflection being satisfied simultaneously and for
each given monochromator are manifested in the strictly determined spectral
positions. In most cases the glitches seen on curves $I_0(E)$ and $I_1(E)$,
vanish on $I_0/I_1$ ratio. If not, one can easily get rid of them. For
instance, the glitch area, usually extremely thin, is smoothed or, with fixed
ends, replaced by a straight-line segment. The main thing in the correct
analysis of glitches is their detection.

\begin{figure}[!t]
\begin{minipage}[c]{0.6\hsize}\includegraphics*{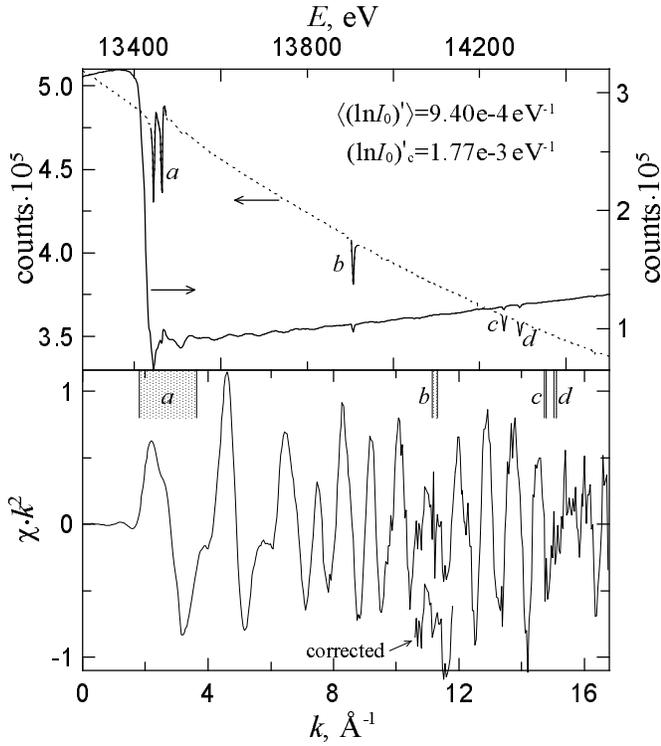}\end{minipage}
\begin{minipage}[c]{0.35\hsize}\caption{Energy dependence of experimental
counts from ion-chambers. Curve $I_0(E)$ relates to the left axis, $I_1(E)$ ---
to the right one. In glitch areas the absolute value of the derivative is
greater than the critical level specified. On $\chi(k)\cdot k^2$ only the
glitch $b$ is manifested. The displaced fragment is $\chi(k)\cdot k^2$ after
correction.} \label{glitch}\end{minipage}\end{figure}

To detect a glitch on curves $\mu$ or $\chi k^w$ is practically impossible.
For this, one needs the primary data $I_0(E)$ and $I_1(E)$, not $\ln(I_0/I_1)$
nor $I_0/I_1$. Out of glitches the intensity of incident radiation smoothly,
ignoring the noise, depends on energy (see Fig.~\ref{glitch}). The idea of
detection of glitches via critical level for the derivative $|d\ln I_0/dE|_c$
is self-evident. For the presented in Fig.~\ref{glitch} $I_0(E)$ curve the
absolute value of the derivative in the glitch areas is greater than the
critical value chosen to be equal to $1.77\cdot10^{-3}$\,í'$^{-1}$. Having
extracted the XAFS, one can see that the first (paired) glitch $a$ is not
manifested on $\chi(k)\cdot k^2$, the last two ($c$ ¨ $d$) are obscured by the
noise, solely glitch $b$ is clearly pronounced. Now, being in the firm belief
that this is not a part of the XAFS, one can eliminate the glitch with ease.
Here, we fixed its ends on $\mu(E)$, replaced it by the straight-line segment,
and constructed $\chi(k)\cdot k^2$ again.

\section{Conclusion}\label{conclusion}
In this paper we have considered all stages of XAFS function extraction from
the measured absorption. We focused our attention on the most important stage,
construction of the atomic-like absorption $\mu_0$.

For the wide-spread method of approximation of $\mu_0$ by a smoothing spline we
have proposed the way to raise the stability by including the prior information
about absorption edge shape (``nearly step'' or ``nearly step with a white
line''). Besides we have propose a new reliable criterion for determination of
the smoothing regularizer.

A new method for approximation of $\mu_0$ is proposed, the method of bayesian
smoothing. It can include various prior information, which raises the accuracy
of XAFS determination. Following this method one finds the distributions of
$\mu_0$ in each experimental point, from which one can find not only average
values but also any desirable momenta, which appears to be an additional
difficulty for other methods. This method was shown to give more accurate
atomic-like background than that obtained by the method of Ref~\cite{Newville1}.

Particular attention has been given to the analysis of noise. We have discussed
the difficulties of its estimates on the basis of statistical approach. More
reliable is the determination of noise from the Fourier transform. We have
shown that the experimental noise is essentially less than the errors of
$\mu_0$ construction, and the use of values of noise in the $\chi^2$-statistics
definition appears to be erroneous since leads to the unjustified optimistic
errors of structural parameters inferred in fitting procedures. For detailed
consideration of the accuracy of fitting parameters see the following paper
\cite{IIe}.

\appendix
\section*{Bayesian smoothing and deconvolution}\label{BayesianMethod}
\subsection{Posterior distribution for smoothed data}

Consider general linear problem of data smoothing with the use of statistical
methods (for introduction see review by Turchin {\em et al.}
\cite{Turchin1eng} and the articles from Web-site {\tt bayes.wustl.edu}). Let
data $\bf d$ are defined on the mesh $x_1,\ldots,x_N$ and consist of the true
values $\bf t$ and the additive noise $\bf n$:
\begin{equation}\label{B1}%%%%%%%%%%%%%%%%%%%%%%%%%%%%%%%%%%%%%%%%%%%%%%%%%%%
d_i=t_i+n_i,\qquad i=1,\ldots,N.
\end{equation}%%%%%%%%%%%%%%%%%%%%%%%%%%%%%%%%%%%%%%%%%%%%%%%%%%%%%%%%%%%%%%%
The problem of smoothing is to find the best estimates of $\bf t$. For an
arbitrary node $j$, find the probability density function for $t_j$ given the
data $\bf d$:
\begin{equation}\label{B2}%%%%%%%%%%%%%%%%%%%%%%%%%%%%%%%%%%%%%%%%%%%%%%%%%%%
P(t_j|{\bf d})=\int\cdots dt_{i\ne j}\cdots P({\bf t}|{\bf d}),
\end{equation}%%%%%%%%%%%%%%%%%%%%%%%%%%%%%%%%%%%%%%%%%%%%%%%%%%%%%%%%%%%%%%%
where $P({\bf t}|{\bf d})$ is the joint probability density function for
all values $\bf t$, and the integration is done over all $t_{i\ne j}$.
According to Bayes theorem,
\begin{equation}\label{B3}%%%%%%%%%%%%%%%%%%%%%%%%%%%%%%%%%%%%%%%%%%%%%%%%%%%
P({\bf t}|{\bf d})=\frac{P({\bf d}|{\bf t})P({\bf t})}{P({\bf d})},
\end{equation}%%%%%%%%%%%%%%%%%%%%%%%%%%%%%%%%%%%%%%%%%%%%%%%%%%%%%%%%%%%%%%%
$P({\bf t})$ being the joint prior probability for all $t_i$, $P({\bf d})$ is
a normalization constant. Assuming that the values $n_i$ are independent in
different nodes and normally distributed with zero expected values, the
probability $P({\bf d}|{\bf t})$, so-called likelihood function, is given by
\begin{equation}\label{B4}%%%%%%%%%%%%%%%%%%%%%%%%%%%%%%%%%%%%%%%%%%%%%%%%%%
P({\bf d}|{\bf t},\sigma)=(2\pi\sigma^2)^{-N/2}
\exp\Bigl(-\frac{1}{2\sigma^2}\sum_{k=1}^{N}(d_k-t_k)^2\Bigr),
\end{equation}%%%%%%%%%%%%%%%%%%%%%%%%%%%%%%%%%%%%%%%%%%%%%%%%%%%%%%%%%%%%%%%
where the standard deviation of the noise, $\sigma$, appears as a known value.
Later, we apply the rules of probability theory to remove $\sigma$ from the
problem.

Now define prior probability $P({\bf t})$. Let we know in advance that the
function $t(x)$ is smooth enough. To specify this information, introduce the
norm of the second derivative and indicate its expected approximate value:
\begin{equation}\label{B5}%%%%%%%%%%%%%%%%%%%%%%%%%%%%%%%%%%%%%%%%%%%%%%%%%%
\Omega(t(x))=\int\left(\frac{d^2t}{dx^2}\right)^2\,dx\approx\omega.
\end{equation}%%%%%%%%%%%%%%%%%%%%%%%%%%%%%%%%%%%%%%%%%%%%%%%%%%%%%%%%%%%%%%
Denote $\Delta_i=x_{i+1}-x_i$, $i=1,\ldots,N-1$ and represent the second
derivative in the finit-difference form:
\begin{equation}\label{B6}%%%%%%%%%%%%%%%%%%%%%%%%%%%%%%%%%%%%%%%%%%%%%%%%
\Omega(t(x))\equiv\Omega({\bf t})=
\sum^{N-1}_{i=2}[t_{i-1}\Delta_{i-1}^{-1}-
t_i(\Delta_{i-1}^{-1}+\Delta_i^{-1})+t_{i+1}\Delta_i^{-1}]^2\equiv
\sum^N_{k,l=1}\Omega_{kl}t_kt_l.
\end{equation}%%%%%%%%%%%%%%%%%%%%%%%%%%%%%%%%%%%%%%%%%%%%%%%%%%%%%%%%%%%%
$\Omega_{kl}$ is a five-diagonal symmetric matrix with the following non-zero
elements:
\begin{eqnarray}\label{B7}%%%%%%%%%%%%%%%%%%%%%%%%%%%%%%%%%%%%%%%%%%%%%%%%%%
\Omega_{11}&=&\Delta_1^{-1}\Delta_2^{-2},\quad
\Omega_{22}=\Delta_2^{-1}(\Delta_1^{-1}+\Delta_2^{-1})^2+\Delta_2^{-2}\Delta_3^{-1},\quad
\Omega_{12}=-(\Delta_1\Delta_2)^{-1}(\Delta_1^{-1}+\Delta_2^{-1}),\\
&&\cdots\cdots\cdots\cdots\nonumber\\
\Omega_{ii}&=&\Delta_i^{-1}(\Delta_i^{-1}+\Delta_{i-1}^{-1})^2+
\Delta_i^{-2}\Delta_{i+1}^{-1}+\Delta_{i-1}^{-3},\nonumber\\
\Omega_{i-1,i}&=&-\Delta_{i-1}^{-2}(\Delta_{i-1}^{-1}+\Delta_{i-2}^{-1})-
(\Delta_{i-1}\Delta_i)^{-1}(\Delta_{i-1}^{-1}+\Delta_i^{-1}),\nonumber\\
\Omega_{i-2,i}&=&\Delta_{i-2}^{-1}\Delta_{i-1}^{-2},\nonumber\\
&&\cdots\cdots\cdots\cdots\nonumber\\
\Omega_{NN}&=&\Delta_{N-1}^{-3},\quad
\Omega_{N-1,N-1}=\Delta_{N-1}^{-1}(\Delta_{N-1}^{-1}+\Delta_{N-2}^{-1})^{2}+
\Delta_{N-2}^{-3},\quad
\Omega_{N-1,N}=-\Delta_{N-1}^{-2}(\Delta_{N-1}^{-1}+\Delta_{N-2}^{-1})\nonumber.
\end{eqnarray}%%%%%%%%%%%%%%%%%%%%%%%%%%%%%%%%%%%%%%%%%%%%%%%%%%%%%%%%%%%%%%
In order to introduce the minimum information in addition to that contained in
(\ref{B6}), from all normalized to unity functions $P({\bf t})$ which satisfy
the condition (\ref{B6}) we choose a single one that contains minimum
information about {\bf t} that is minimizes the functional
\begin{equation}\label{B8}%%%%%%%%%%%%%%%%%%%%%%%%%%%%%%%%%%%%%%%%%%%%%%%%%%
I[P({\bf t})]=\int P({\bf t})\ln P({\bf t})\,d{\bf t}+
\beta\Bigl[1-\int P({\bf t})\,d{\bf t}\Bigr]+
\gamma\Bigl[\omega-\int\Omega({\bf t})\,P({\bf t})\,d{\bf t}\Bigr],
\end{equation}%%%%%%%%%%%%%%%%%%%%%%%%%%%%%%%%%%%%%%%%%%%%%%%%%%%%%%%%%%%%%%
where $\beta$ and $\gamma$ are the Larrange multipliers. In minimizing
$I[P({\bf t})]$, one obtains the equation set
\begin{eqnarray}\label{B9}%%%%%%%%%%%%%%%%%%%%%%%%%%%%%%%%%%%%%%%%%%%%%%%%%%
\ln P({\bf t})+1-\beta-\gamma\Omega({\bf t})&=&0\\
\int P({\bf t})\,d{\bf t}&=&1\nonumber\\
\int\Omega({\bf t})\,P({\bf t})\,d{\bf t}&=&\omega\nonumber,
\end{eqnarray}%%%%%%%%%%%%%%%%%%%%%%%%%%%%%%%%%%%%%%%%%%%%%%%%%%%%%%%%%%%%%%
that has a solution:
\begin{equation}\label{B10}%%%%%%%%%%%%%%%%%%%%%%%%%%%%%%%%%%%%%%%%%%%%%%%%%%
P({\bf t})=(\lambda_1\cdots\lambda_N)^{-1/2}
\Bigl(\frac{2\pi\sigma^2}{\alpha}\Bigr)^{-N/2}
\exp\Bigl(-\frac{\alpha}{2\sigma^2}\Omega({\bf t})\Bigr),
\end{equation}%%%%%%%%%%%%%%%%%%%%%%%%%%%%%%%%%%%%%%%%%%%%%%%%%%%%%%%%%%%%%%
where $\alpha/2\sigma^2=\gamma=N/2\omega$, and $\lambda_1,\ldots,\lambda_N$ are
the eigenvalues of the matrix $\Omega_{kl}$. The regularizer $\alpha$
will be used to control the smoothness of ${\bf t}$. The prior
distribution obtained is a ``soft'' one, that is does not demand from the
solution to have a strictly prescribed form.

Thus, we have for the probability density function:
\begin{eqnarray}\label{B11}%%%%%%%%%%%%%%%%%%%%%%%%%%%%%%%%%%%%%%%%%%%%%%%%%%%
P(t_j|{\bf d},\sigma,\alpha)&\propto&\int\cdots dt_{i\ne j}\cdots
\sigma^{-2N}\alpha^{N/2}
\exp\Bigl(-\frac{\alpha}{2\sigma^2}\sum^N_{k,l=1}\Omega_{kl}t_kt_l\Bigr)
\exp\Bigl(-\frac{1}{2\sigma^2}\sum_{k=1}^{N}(d_k-t_k)^2\Bigr)\nonumber\\
&=&\int\cdots dt_{i\ne j}\cdots \sigma^{-2N}\alpha^{N/2}
\exp\Bigl(-\frac{1}{2\sigma^2}\bigl[{\bf d}^2-2\sum^{N}_{k=1}d_kt_k+
\sum_{k,l=1}^{N}g_{kl}t_kt_l\bigr]\Bigr),
\end{eqnarray}%%%%%%%%%%%%%%%%%%%%%%%%%%%%%%%%%%%%%%%%%%%%%%%%%%%%%%%%%%%%%%%
where
\begin{eqnarray}\label{B11_}%%%%%%%%%%%%%%%%%%%%%%%%%%%%%%%%%%%%%%%%%%%%%%%%%%%
g_{kl}=\alpha\Omega_{kl}+\delta_{kl},\qquad{\bf d}^2=\sum_{k=1}^N d_k^2.
\end{eqnarray}%%%%%%%%%%%%%%%%%%%%%%%%%%%%%%%%%%%%%%%%%%%%%%%%%%%%%%%%%%%%%%%
Since there is no integral over $t_j$, separate it from the other integration
variables:
\begin{eqnarray}\label{B12}%%%%%%%%%%%%%%%%%%%%%%%%%%%%%%%%%%%%%%%%%%%%%%%%%%%
P(t_j|{\bf d},\sigma,\alpha)&\propto&\sigma^{-2N}\alpha^{N/2}
\exp\Bigl(-\frac{1}{2\sigma^2}[{\bf d}^2-2d_jt_j+g_{jj}t_j^2]\Bigr)\nonumber\\
&&\times\int\cdots dt_{i\ne j}\cdots
\exp\Bigl(-\frac{1}{2\sigma^2}\bigl[\mathop{{\sum}\smash{^j}}_{k,l=1}^{N}g_{kl}t_kt_l-
2\mathop{{\sum}\smash{^j}}_{k=1}^{N}[d_k-g_{kj}t_j]t_k\bigr]\Bigr),
\end{eqnarray}%%%%%%%%%%%%%%%%%%%%%%%%%%%%%%%%%%%%%%%%%%%%%%%%%%%%%%%%%%%%%%%
Here, the symbol $j$ near the summation signs denotes the absence of $j$-th
item. Further, find the eigenvalues $\lambda'_i$ and corresponding eigenvectors
${\bf e}_i$ of the matrix $g_{kl}$ in which the $j$-th row and column are
deleted, and change the variables:
\begin{equation}\label{B13}%%%%%%%%%%%%%%%%%%%%%%%%%%%%%%%%%%%%%%%%%%%%%%%%%%
b_i=\sqrt{\lambda'_i}\mathop{{\sum}\smash{^j}}_{k=1}^{N}t_ke_{ik},\qquad
t_k=\mathop{{\sum}\smash{^j}}_{i=1}^{N}\frac{b_ie_{ik}}{\sqrt{\lambda'_i}}
\qquad (i,k\ne j).
\end{equation}%%%%%%%%%%%%%%%%%%%%%%%%%%%%%%%%%%%%%%%%%%%%%%%%%%%%%%%%%%%%%%
Using the properties of eigenvectors:
\begin{equation}\label{B14}%%%%%%%%%%%%%%%%%%%%%%%%%%%%%%%%%%%%%%%%%%%%%%%%%%
\mathop{{\sum}\smash{^j}}_{k=1}^{N}g_{lk}e_{ik}=\lambda'_ie_{il},\qquad
\mathop{{\sum}\smash{^j}}_{k=1}^{N}e_{lk}e_{ik}=\delta_{li}\qquad (l,i\ne j),
\end{equation}%%%%%%%%%%%%%%%%%%%%%%%%%%%%%%%%%%%%%%%%%%%%%%%%%%%%%%%%%%%%%%
one obtains:
\begin{eqnarray}\label{B15}%%%%%%%%%%%%%%%%%%%%%%%%%%%%%%%%%%%%%%%%%%%%%%%%%%%
P(t_j|{\bf d},\sigma,\alpha)&\propto&\sigma^{-2N}\alpha^{N/2}
\exp\Bigl(-\frac{1}{2\sigma^2}[({\bf d}^2-{\bf h}^2)-2t_j(d_j-{\bf h}{\bf u})+
t_j^2(g_{jj}-{\bf u}^2)]\Bigr)\nonumber\\
&&\times\int\cdots db_{l\ne j}\cdots
\exp\Bigl(-\frac{1}{2\sigma^2}\mathop{{\sum}\smash{^j}}_{i=1}^{N}
[b_i-h_i+u_it_j]^2\Bigr),
\end{eqnarray}%%%%%%%%%%%%%%%%%%%%%%%%%%%%%%%%%%%%%%%%%%%%%%%%%%%%%%%%%%%%%%%
where new quantities were introduced:
\begin{eqnarray}\label{B16}%%%%%%%%%%%%%%%%%%%%%%%%%%%%%%%%%%%%%%%%%%%%%%%%%%%
h_i=\frac{1}{\sqrt{\lambda'_i}}\mathop{{\sum}\smash{^j}}_{k=1}^{N}d_ke_{ik},\qquad
u_i=\frac{1}{\sqrt{\lambda'_i}}\mathop{{\sum}\smash{^j}}_{k=1}^{N}g_{kj}e_{ik},\qquad
\nonumber\\
{\bf h}^2=\mathop{{\sum}\smash{^j}}_{i=1}^{N}h_i^2,\qquad
{\bf u}^2=\mathop{{\sum}\smash{^j}}_{i=1}^{N}u_i^2,\qquad
{\bf h}{\bf u}=\mathop{{\sum}\smash{^j}}_{i=1}^{N}h_iu_i.
\end{eqnarray}%%%%%%%%%%%%%%%%%%%%%%%%%%%%%%%%%%%%%%%%%%%%%%%%%%%%%%%%%%%%%%%
Evaluating the $N-1$ integrals in (\ref{B15}), one finally obtains the
posterior probability for $j$-th node:
\begin{equation}\label{B17}%%%%%%%%%%%%%%%%%%%%%%%%%%%%%%%%%%%%%%%%%%%%%%%%%%%
P(t_j|{\bf d},\sigma,\alpha)\propto\sigma^{-(N+1)}\alpha^{N/2}
\exp\Bigl(-\frac{1}{2\sigma^2}
[({\bf d}^2-{\bf h}^2)-2t_j(d_j-{\bf h}{\bf u})+t_j^2(g_{jj}-{\bf u}^2)]\Bigr).
\end{equation}%%%%%%%%%%%%%%%%%%%%%%%%%%%%%%%%%%%%%%%%%%%%%%%%%%%%%%%%%%%%%%%

\subsection{Eliminating nuisance parameters}
In most real problems $\sigma$ and $\alpha$ are not known. To eliminate
$\sigma$ is a quite straightforward problem:
\begin{equation}\label{B18}%%%%%%%%%%%%%%%%%%%%%%%%%%%%%%%%%%%%%%%%%%%%%%%%%%%
P(t_j|{\bf d},\alpha)=\int d\sigma P(t_j,\sigma|{\bf d},\alpha)=
\int d\sigma P(\sigma)P(t_j|{\bf d},\sigma,\alpha),
\end{equation}%%%%%%%%%%%%%%%%%%%%%%%%%%%%%%%%%%%%%%%%%%%%%%%%%%%%%%%%%%%%%%%
one needs only to know a prior probability $P(\sigma)$. Having no specific
information about $\sigma$, a Jeffreys prior $P(\sigma)=1/\sigma$ is assigned
\cite{Jeffreys1}. Then
\begin{eqnarray}\label{B19}%%%%%%%%%%%%%%%%%%%%%%%%%%%%%%%%%%%%%%%%%%%%%%%%%%%
P(t_j|{\bf d},\alpha)&\propto&\int_0^\infty d\sigma \sigma^{-(N+2)}
\exp\Bigl(-\frac{1}{2\sigma^2}
[({\bf d}^2-{\bf h}^2)-2t_j(d_j-{\bf h}{\bf u})+t_j^2(g_{jj}-{\bf u}^2)]\Bigr)\\
&\propto&
[({\bf d}^2-{\bf h}^2)-2t_j(d_j-{\bf h}{\bf u})+t_j^2(g_{jj}-{\bf
u}^2)]^{-(N+1)/2} \nonumber.
\end{eqnarray}%%%%%%%%%%%%%%%%%%%%%%%%%%%%%%%%%%%%%%%%%%%%%%%%%%%%%%%%%%%%%%%
Introducing the substitution
\begin{eqnarray}\label{B20}%%%%%%%%%%%%%%%%%%%%%%%%%%%%%%%%%%%%%%%%%%%%%%%%%%%
w_j^2=N\frac{(g_{jj}-{\bf u}^2)^2}{({\bf d}^2-{\bf h}^2)(g_{jj}-{\bf u}^2)-
(d_j-{\bf h}{\bf u})^2}\left(t_j-\frac{d_j-{\bf h}{\bf u}}{g_{jj}-{\bf
u}^2}\right)^2,
\end{eqnarray}%%%%%%%%%%%%%%%%%%%%%%%%%%%%%%%%%%%%%%%%%%%%%%%%%%%%%%%%%%%%%%%
one obtains the Student $t$-distribution with $N$ degrees of freedom:
\begin{eqnarray}\label{B21}%%%%%%%%%%%%%%%%%%%%%%%%%%%%%%%%%%%%%%%%%%%%%%%%%%%
P(w_j|{\bf d},\alpha)\propto\Bigl(1+\frac{w_j^2}{N}\Bigr)^{-(N+1)/2}
\end{eqnarray}%%%%%%%%%%%%%%%%%%%%%%%%%%%%%%%%%%%%%%%%%%%%%%%%%%%%%%%%%%%%%%%
with zero average and the variance $N/(N-2)$. From where one finds for $t_j$:
\begin{eqnarray}\label{B22}%%%%%%%%%%%%%%%%%%%%%%%%%%%%%%%%%%%%%%%%%%%%%%%%%%%
\bar t_j=\frac{d_j-{\bf h}{\bf u}}{g_{jj}-{\bf u}^2},\qquad\
\delta^2(t_j)=\frac{({\bf d}^2-{\bf h}^2)(g_{jj}-{\bf u}^2)-
(d_j-{\bf h}{\bf u})^2}{(g_{jj}-{\bf u}^2)^2}\frac{1}{N-2}.
\end{eqnarray}%%%%%%%%%%%%%%%%%%%%%%%%%%%%%%%%%%%%%%%%%%%%%%%%%%%%%%%%%%%%%%%
Thus, we have got rid of unknown $\sigma$ and found the expressions for mean
values $t_j$ and their dispersions at known regularizer $\alpha$. To eliminate
the latter is more difficult. The idea is not to find the smoothest solution,
but the solution of the most probable smoothness. For that we will find the
posterior probability:
\begin{eqnarray}\label{B23}%%%%%%%%%%%%%%%%%%%%%%%%%%%%%%%%%%%%%%%%%%%%%%%%%%%
P(\alpha|{\bf d})=\int d{\bf t}d\sigma P(\alpha,\sigma,{\bf t}|{\bf d})=
\int d{\bf t}d\sigma P(\alpha,\sigma)P({\bf t}|\alpha,\sigma,{\bf d}).
\end{eqnarray}%%%%%%%%%%%%%%%%%%%%%%%%%%%%%%%%%%%%%%%%%%%%%%%%%%%%%%%%%%%%%%%
Assuming that $\alpha$ and $\sigma$ are independent and using Bayes theorem
(\ref{B3}), one obtains:
\begin{eqnarray}\label{B24}%%%%%%%%%%%%%%%%%%%%%%%%%%%%%%%%%%%%%%%%%%%%%%%%%%%
P(\alpha|{\bf d})\propto\int d{\bf t}d\sigma P(\alpha)P(\sigma)
P({\bf t}|\alpha,\sigma)P({\bf d}|{\bf t},\alpha,\sigma).
\end{eqnarray}%%%%%%%%%%%%%%%%%%%%%%%%%%%%%%%%%%%%%%%%%%%%%%%%%%%%%%%%%%%%%%%
Substituting (\ref{B10}) for the prior probability $P({\bf t}|\alpha,\sigma)$,
(\ref{B4}) for the likelihood, and a Jeffreys prior $P(\sigma)=1/\sigma$ and
$P(\alpha)=1/\alpha$, one obtains the posterior distribution for the
regularizer $\alpha$:
\begin{eqnarray}\label{B25}%%%%%%%%%%%%%%%%%%%%%%%%%%%%%%%%%%%%%%%%%%%%%%%%%%%
P(\alpha|{\bf d})&\propto&\int d{\bf t}d\sigma\sigma^{-2N-1}\alpha^{N/2-1}
\exp\Bigl(-\frac{\alpha}{2\sigma^2}\sum^N_{k,l=1}\Omega_{kl}t_kt_l\Bigr)
\exp\Bigl(-\frac{1}{2\sigma^2}\sum_{k=1}^{N}(d_k-t_k)^2\Bigr)\nonumber\\
&=&\int d{\bf t}d\sigma\sigma^{-2N-1}\alpha^{N/2-1}
\exp\Bigl(-\frac{1}{2\sigma^2}\bigl[{\bf d}^2-2\sum^{N}_{k=1}d_kt_k+
\sum_{k,l=1}^{N}g_{kl}t_kt_l\bigr]\Bigr),
\end{eqnarray}%%%%%%%%%%%%%%%%%%%%%%%%%%%%%%%%%%%%%%%%%%%%%%%%%%%%%%%%%%%%%%%
where matrix $g_{kl}$ was defined in (\ref{B11_}). After its diagonalization,
analogously to what was done above, finally one obtains:
\begin{eqnarray}\label{B26}%%%%%%%%%%%%%%%%%%%%%%%%%%%%%%%%%%%%%%%%%%%%%%%%%%%
P(\alpha|{\bf d})\propto (\lambda'_1\cdots\lambda'_N)^{-1/2}\alpha^{N/2-1}
[{\bf d}^2-{\bf h}^2]^{-N/2},
\end{eqnarray}%%%%%%%%%%%%%%%%%%%%%%%%%%%%%%%%%%%%%%%%%%%%%%%%%%%%%%%%%%%%%%%
where ${\bf h}^2$ is given by
\begin{eqnarray}\label{B16_1}%%%%%%%%%%%%%%%%%%%%%%%%%%%%%%%%%%%%%%%%%%%%%%%%%%%
{\bf h}^2=\sum_{i=1}^{N}h_i^2,\qquad
h_i=\frac{1}{\sqrt{\lambda'_i}}\sum_{k=1}^{N}d_ke_{ik},
\end{eqnarray}%%%%%%%%%%%%%%%%%%%%%%%%%%%%%%%%%%%%%%%%%%%%%%%%%%%%%%%%%%%%%%%
and $\lambda'_i$ and ${\bf e}_i$ are the eigenvalues and eigenvectors of
$g_{kl}$. Having found the maximum of the posterior probability (\ref{B26})
or having averaged over it the expression (\ref{B22}), one has the sought
${\bf t}$ with the most probable smoothness. However it is necessary to point
out that this procedure narrows the applicability of the bayesian smoothing
down to the class of tasks where the smoothed values lie in most within the
limits $\pm\sigma$ from the most probable. In practice, there possible other
tasks where the condition (\ref{B1}) is treated more wider and the smoothed
values exceed the bounds of noise.

\subsection{Expressions for smoothed values and their variances}
The formulas (\ref{B22}) appear useless in practice since require to find the
eigenvalues and eigenvectors for the matrix of rank $N-1$ on each node. Those
formulas have merely a methodological value: the explicit expressions for
posterior probabilities enable one to find the average of {\em arbitrary}
function of $t_j$. However, $\bar t_j$ and $\delta^2(t_j)$ could be found
significantly easier. Using (\ref{B18}) and (\ref{B11}), represent $\bar t_j$
as:
\begin{eqnarray}\label{B26_1}%%%%%%%%%%%%%%%%%%%%%%%%%%%%%%%%%%%%%%%%%%%%%%%%%%% \bar t_j&=&\int t_jP(t_j|{\bf d},\alpha)dt_j=\int t_jd\sigma
P(\sigma)P(t_j|{\bf d},\sigma,\alpha)dt_j\nonumber\\
&\propto&\int d{\bf t}d\sigma \sigma^{-2N-1}t_j
\exp\Bigl(-\frac{1}{2\sigma^2}\bigl[{\bf d}^2-2\sum^{N}_{k=1}d_kt_k+
\sum_{k,l=1}^{N}g_{kl}t_kt_l\bigr]\Bigr).
\end{eqnarray}%%%%%%%%%%%%%%%%%%%%%%%%%%%%%%%%%%%%%%%%%%%%%%%%%%%%%%%%%%%%%%%
Performing the diagonalization, one obtains:
\begin{eqnarray}\label{B26_2}%%%%%%%%%%%%%%%%%%%%%%%%%%%%%%%%%%%%%%%%%%%%%%%%%%%
\bar t_j&\propto&\int d{\bf b}d\sigma \sigma^{-2N-1}
\exp\Bigl(-\frac{1}{2\sigma^2}[{\bf d}^2-{\bf h}^2]\Bigr)
\Bigl(\sum_{i=1}^{N}\frac{b_ie_{ij}}{\sqrt{\lambda'_i}}\Bigr)
\exp\Bigl(-\frac{1}{2\sigma^2}\sum_{i=1}^{N}[b_i-h_i]^2\Bigr)\nonumber\\
&\propto&\sum_{i=1}^{N}\frac{h_ie_{ij}}{\sqrt{\lambda'_i}}
\int d\sigma \sigma^{-N-1}
\exp\Bigl(-\frac{1}{2\sigma^2}[{\bf d}^2-{\bf h}^2]\Bigr),
\end{eqnarray}%%%%%%%%%%%%%%%%%%%%%%%%%%%%%%%%%%%%%%%%%%%%%%%%%%%%%%%%%%%%%%%
from where
\begin{eqnarray}\label{B26_3}%%%%%%%%%%%%%%%%%%%%%%%%%%%%%%%%%%%%%%%%%%%%%%%%%%%
\bar t_j=\sum_{i=1}^{N}\frac{h_ie_{ij}}{\sqrt{\lambda'_i}}.
\end{eqnarray}%%%%%%%%%%%%%%%%%%%%%%%%%%%%%%%%%%%%%%%%%%%%%%%%%%%%%%%%%%%%%%%
Analogously, for the variance $\delta(t_j)$ one has:
\begin{eqnarray}\label{B26_4}%%%%%%%%%%%%%%%%%%%%%%%%%%%%%%%%%%%%%%%%%%%%%%%%%%%
\delta^2(t_j)&\propto&\int d{\bf b}d\sigma \sigma^{-2N-1}
\exp\Bigl(-\frac{1}{2\sigma^2}[{\bf d}^2-{\bf h}^2]\Bigr)
\Bigl(\sum_{i=1}^{N}\frac{(b_i-h_i)e_{ij}}{\sqrt{\lambda'_i}}\Bigr)^2
\exp\Bigl(-\frac{1}{2\sigma^2}\sum_{i=1}^{N}[b_i-h_i]^2\Bigr)\nonumber\\
&\propto&\sum_{i=1}^{N}\frac{e_{ij}^2}{\lambda'_i}
\int d\sigma \sigma^{-N-1}
\exp\Bigl(-\frac{1}{2\sigma^2}[{\bf d}^2-{\bf h}^2]\Bigr)\sigma^2.
\end{eqnarray}%%%%%%%%%%%%%%%%%%%%%%%%%%%%%%%%%%%%%%%%%%%%%%%%%%%%%%%%%%%%%%%
Normalizing, one finally obtains:
\begin{eqnarray}\label{B26_5}%%%%%%%%%%%%%%%%%%%%%%%%%%%%%%%%%%%%%%%%%%%%%%%%%%%
\delta^2(t_j)&=&\frac{\int d\sigma \sigma^{-N+1}
\exp\left(-[{\bf d}^2-{\bf h}^2]/2\sigma^2\right)}
{\int d\sigma \sigma^{-N-1}
\exp\left(-[{\bf d}^2-{\bf h}^2]/2\sigma^2\right)}
\sum_{i=1}^{N}\frac{e_{ij}^2}{\lambda'_i}\nonumber\\
&=&\ \frac{\Gamma\!\left(\frac{N}{2}-1\right)}
{\left([{\bf d}^2-{\bf h}^2]/2\right)^{N/2-1}}\
\frac{\left([{\bf d}^2-{\bf h}^2]/2\right)^{N/2}}
{\Gamma\!\left(\frac{N}{2}\right)}\sum_{i=1}^{N}\frac{e_{ij}^2}{\lambda'_i}=
\frac{[{\bf d}^2-{\bf h}^2]}{N-2}\sum_{i=1}^{N}\frac{e_{ij}^2}{\lambda'_i}.
\end{eqnarray}%%%%%%%%%%%%%%%%%%%%%%%%%%%%%%%%%%%%%%%%%%%%%%%%%%%%%%%%%%%%%%%
Now we got the usable formulas, which require to find the eigenvalues and
eigenvectors for the matrix of rank $N$ just {\em one time}.

\subsection{Addenda to the bayesian smoothing}
(i) Let the curvature of the function $t(x)$ is approximately known in advance.
To specify this information, introduce the norm of the difference between
$d^2t/dx^2$ and approximately known second derivative $d^2\!f/dx^2$:
\begin{equation}\label{B27}%%%%%%%%%%%%%%%%%%%%%%%%%%%%%%%%%%%%%%%%%%%%%%%%%%
\Omega(t(x))=\int\left(\frac{d^2t}{dx^2}-\frac{d^2\!f}{dx^2}\right)^2\,
dx\approx\omega.
\end{equation}%%%%%%%%%%%%%%%%%%%%%%%%%%%%%%%%%%%%%%%%%%%%%%%%%%%%%%%%%%%%%%
Notice, that there is no need to know $f(x)$ itself, its second derivative is
sufficient. The explicit presence of $f(x)$ in the following formulas should be
taken as a consequence of the technical trick applied: at first $f(x)$ is
subtracted from the data, then it is added to the found solution.

Everywhere in formulas (\ref{B6}--\ref{B26_5}) make the substitutions:
\begin{equation}\label{B28}%%%%%%%%%%%%%%%%%%%%%%%%%%%%%%%%%%%%%%%%%%%%%%%%%%%
\tilde t_i=t_i-f_i,\qquad\tilde d_i=d_i-f_i,\qquad i=1,\ldots,N.
\end{equation}%%%%%%%%%%%%%%%%%%%%%%%%%%%%%%%%%%%%%%%%%%%%%%%%%%%%%%%%%%%%%%%
Performing the described above procedure for smoothing, one finds $\tilde t_i$,
from which by inverse transformation the sought vector is given by
${\bf t}=\tilde{{\bf t}}+{\bf f}$.

(ii) In some tasks the value on the starting (zero) node is known without
measurement. This sort of prior information represents a ``hard'' one, that is
it restricted the class of possible solutions; in the given case the solution
must pass through the known zero node. The quadratic form $\Omega({\bf t})$
(or $\Omega(\tilde{{\bf t}})$ in the case of approximately known second
derivative) in the expression for the prior probability has changed:
\begin{equation}\label{B29}%%%%%%%%%%%%%%%%%%%%%%%%%%%%%%%%%%%%%%%%%%%%%%%%
\Omega({\bf t})=\sum^{N-1}_{i=1}[t_{i-1}\Delta_{i-1}^{-1}-
t_i(\Delta_{i-1}^{-1}+\Delta_i^{-1})+t_{i+1}\Delta_i^{-1}]^2\equiv
\sum^N_{k,l=1}\Omega_{kl}t_kt_l+
\Omega_{00}t_0^2+2\Omega_{01}t_0t_1+2\Omega_{02}t_0t_2,
\end{equation}%%%%%%%%%%%%%%%%%%%%%%%%%%%%%%%%%%%%%%%%%%%%%%%%%%%%%%%%%%%%
the first few matrix elements of $\Omega_{kl}$ now are:
\begin{eqnarray}\label{B30}%%%%%%%%%%%%%%%%%%%%%%%%%%%%%%%%%%%%%%%%%%%%%%%%%%
\Omega_{00}&=&\Delta_0^{-2}\Delta_1^{-1},\quad
\Omega_{01}=-(\Delta_0\Delta_1)^{-1}(\Delta_0^{-1}+\Delta_1^{-1}),\quad
\Omega_{02}=\Delta_0^{-1}\Delta_1^{-2},\\
\Omega_{11}&=&\Delta_1^{-1}(\Delta_0^{-1}+\Delta_1^{-1})^2+\Delta_1^{-1}\Delta_2^{-2},\quad
\Omega_{12}=-\Delta_1^{-2}(\Delta_1^{-1}+\Delta_0^{-1})-
(\Delta_1\Delta_2)^{-1}(\Delta_1^{-1}+\Delta_2^{-1})\nonumber.
\end{eqnarray}%%%%%%%%%%%%%%%%%%%%%%%%%%%%%%%%%%%%%%%%%%%%%%%%%%%%%%%%%%%%%%
If $t_0=0$ (or $\tilde t_0=0$), none further changes to the formulas of
smoothing (\ref{B6}--\ref{B26_5}) are needed; at $t_0\ne0$ the changes are
evident: instead of the scalar product ${\bf d}{\bf t}$ in (\ref{B11}) will be
$\rm(\bf d-\bf\hat d)\bf t$, where $\hat d_1=\alpha t_0\Omega_{01}$,
$\hat d_2=\alpha t_0\Omega_{02}$, all remaining $\hat d_i=0$; to the ${\bf d}^2$
the term $\alpha t_0^2\Omega_{00}$ will be added.

(iii) Making some changes in the considered above problem of smoothing allows
one to solve the problem of deconvolution. If the experimental value $d_j$ on
some node $j$ is determined not only by $t_j$ but also by the values of some
neighboring nodes, then instead of (\ref{B1}) we have:
\begin{eqnarray}\label{B31}%%%%%%%%%%%%%%%%%%%%%%%%%%%%%%%%%%%%%%%%%%%%%%%%%%
d_i=\sum^{N}_{j=1}r_{ij}t_j+n_i,\qquad i=1,\ldots,N,
\end{eqnarray}%%%%%%%%%%%%%%%%%%%%%%%%%%%%%%%%%%%%%%%%%%%%%%%%%%%%%%%%%%%%%%
where $r_{ij}$ is the grid representation of the impulse response function.
Instead of expression (\ref{B4}), for the likelihood now we have:
\begin{equation}\label{B32}%%%%%%%%%%%%%%%%%%%%%%%%%%%%%%%%%%%%%%%%%%%%%%%%%%
P({\bf d}|{\bf t},\sigma)=(2\pi\sigma^2)^{-N/2}\exp\Bigl(-\frac{1}{2\sigma^2}
\sum_{k=1}^{N}\Bigl[d_k-\sum_{i=1}^{N}r_{ik}t_i\Bigr]^2\Bigr),
\end{equation}%%%%%%%%%%%%%%%%%%%%%%%%%%%%%%%%%%%%%%%%%%%%%%%%%%%%%%%%%%%%%%%
and instead of (\ref{B11}), the posterior probability for $t_j$ is now
expressed as:
\begin{eqnarray}\label{B33}%%%%%%%%%%%%%%%%%%%%%%%%%%%%%%%%%%%%%%%%%%%%%%%%%%%
P(t_j|{\bf d},\sigma,\alpha)\propto\int\cdots dt_{i\ne j}\cdots
\sigma^{-2N}\alpha^{N/2}
\exp\Bigl(-\frac{1}{2\sigma^2}\bigl[{\bf d}^2-2\sum^{N}_{k=1}D_kt_k+
\sum_{k,l=1}^{N}G_{kl}t_kt_l\bigr]\Bigr),
\end{eqnarray}%%%%%%%%%%%%%%%%%%%%%%%%%%%%%%%%%%%%%%%%%%%%%%%%%%%%%%%%%%%%%%%
where
\begin{eqnarray}\label{B34}%%%%%%%%%%%%%%%%%%%%%%%%%%%%%%%%%%%%%%%%%%%%%%%%%%%
G_{kl}=\alpha\Omega_{kl}+\sum^{N}_{i=1}r_{ik}r_{il},\quad
D_k=\sum^{N}_{i=1}r_{ik}d_i.
\end{eqnarray}%%%%%%%%%%%%%%%%%%%%%%%%%%%%%%%%%%%%%%%%%%%%%%%%%%%%%%%%%%%%%%%
Further steps for finding of ${\bf t}$ are analogous to the described above.

%\bibliography{EXAFS,bayes}
%\bibliographystyle{my_prsty}

\end{document}